\documentstyle[psfig]{mn}

\def\gs{\mathrel{\raise0.35ex\hbox{$\scriptstyle >$}\kern-0.6em
\lower0.40ex\hbox{{$\scriptstyle \sim$}}}}
\def\ls{\mathrel{\raise0.35ex\hbox{$\scriptstyle <$}\kern-0.6em
\lower0.40ex\hbox{{$\scriptstyle \sim$}}}}
\def\ls{\mathrel{\hbox{\rlap{\hbox{\lower4pt\hbox{$\sim$}}}\hbox{$<$}}}}
\def\gs{\mathrel{\hbox{\rlap{\hbox{\lower4pt\hbox{$\sim$}}}\hbox{$>$}}}}

\title[Detecting Galaxy Filaments]
      {A New Algorithm for the Detection of Inter-cluster Galaxy Filaments
using Galaxy Orientation Alignments}
\author[K.\,A.\ Pimbblet]
       {Kevin A.\ Pimbblet
        \vspace*{1mm}\\
        Department of Physics, University of Queensland, Brisbane,
        4072 Queensland, Australia\\
        pimbblet@physics.uq.edu.au}

\date{\fbox{\sc Draft: \today\ --- Do Not Distribute}}
\pagerange{000--000}

\begin{document}

\maketitle

\begin{abstract}
We present a new algorithm to detect inter-cluster galaxy filaments
based upon the assumption that the orientations of constituent galaxies 
along such filaments are non-isotropic.  We apply the algorithm to
the 2dF Galaxy Redshift Survey catalogue and find that it readily 
detects many straight filaments between close cluster pairs.  At large 
inter-cluster separations ($>>15$h$^{\rm -1}$ Mpc), 
we find that the detection efficiency falls quickly, as it also does 
with more complex filament morphologies.  We explore the underlying 
assumptions and suggest that it is only in the case of close cluster
pairs that we can expect galaxy orientations to be significantly
correlated with filament direction.
\end{abstract}


\section{Introduction}

Hierarchical structure formation models have long
predicted that galaxy clusters grow through 
repeated mergers with other galaxy clusters and galaxy groups
and continuous accretion of their surrounding matter 
(e.g.\ Zeldovich, Einasto \& Shandarin 1982; Katz et al.\ 1996;
Jenkins et al.\ 1998; Colberg et al.\ 2000;
see also Bond, Kofman \& Pogosyan 1996).
Moreover, the accretion process usually happens in a very
non-isotropic manner: galaxy filaments funnel matter onto 
large clusters along preferred directions (see Ebeling, Barrett 
\& Donovan 2004; Kodama et al.\ 2001).
Beyond the cluster core (say $>$ few virial radii), 
galaxy filaments are predicted (Colberg, Krughoff \& Connolly
2004 and references therein) and observed to inter-connect 
many galaxy clusters in a complex, web-like manner
(Pimbblet, Drinkwater \& Hawkrigg 2004; Pimbblet \& Drinkwater 2004;
Dietrich et al.\ 2004; Gal \& Lubin 2004; Ebeling, Barrett \& 
Donovan 2004; Durret et al.\ 2003; Arnaud et al.\ 2000; Scharf et al.\ 2000;
Kull \& Bohringer 1999; Connolly et al.\ 1996 amongst others).
It is this cosmic web that gives modern redshift surveys
their striking \& characteristic visual appearance (LCRS; 2dFGRS; SDSS)
and quantifying the web's galaxy distribution 
and large-scale morphology has been a central
focus of modern cosmology (e.g.\ 
Pandey \& Bharadwaj 2004; Bharadwaj et al. 2004; 
Sheth 2004; Hikage et al.\ 2002;
Hoyle at al.\ 2002; Tegmark et al.\ 2002; Connolly et al.\ 2002;
Szapudi et al.\ 2002; Sahni et al.\ 1998; 
Shandarin \& Yess 1998; Mecke, Buchert \& Wagner 1994;
Maddox et al.\ 1990;
Gott, Dickinson \& Mellott 1986; Peebles \& Groth 1975).

Filaments of galaxies (FOGs herein) are known to be highly important
for the mass budget of the Universe (e.g.\ Colberg et al.\ 1999).
Indeed, Cen \& Ostriker (1999) show that for a $\Lambda$ cold dark
matter ($\Lambda$CDM) Universe, a large fraction, perhaps as
much as half (Fukugita, Hogan \& Peebles 1998), 
of baryonic material will not have been observed
as it is situated in the inter-cluster media in a hot and
tenuous gaseous phase.  Along with the dark matter component
and perhaps up to a quarter of the galaxian population, these
baryons are preferentially situated in (inter-cluster) FOGs.  
Moreover, FOGs can provide tests of structure formation
(cf.\ Colberg, Krughoff \& Connolly 2004 with Pimbblet, Drinkwater
\& Hawkrigg 2004) and cluster evolution (see Colberg et al.\ 1999).
Indeed, it is precisely in this low density 
filamentary regime that suppression of star-formation rates 
begins to occur (e.g.\ Balogh et al.\ 2004; Gomez et al.\ 2003; 
Lewis et al.\ 2002; see also Pimbblet 2003).  

Finding FOGs is therefore becoming an important task in 
examining both structure formation theory and 
the evolution of stellar populations in galaxy clusters, yet there
remains no good, single method to detect them.
One method to find them was piloted by
Briel \& Henry (1995) who attempted to find evidence of 
the hot inter-cluster gas by searching for X-ray emission
(from thermal bremsstrahlung) between galaxy clusters
using {\it ROSAT} All-Sky Survey data.
Although unsuccessful, the search yielded an 
X-ray surface brightness upper bound
of $4\times10^{-16}$ ergs cm$^{\rm{-2}}$ s$^{\rm{-1}}$ (0.5 --
2.0 keV) on any inter-cluster FOG present in their sample.
Scharf et al.\ (2000) make a $5\sigma$ 
joint X-ray/optical detection of $>12 h^{-1}_{50}$ Mpc
(0.5 deg) FOG with a surface brightness of
$1.6\times10^{-16}$ ergs cm$^{\rm{-2}}$ s$^{\rm{-1}}$.
In the Shapley supercluster meanwhile, Kull \& Bohringer (1999)
find extended X-ray emission between a close cluster pair
that is $\sim 2.5$ times brighter than Briel \& Henry's (1995)
bound, but this could be due to the clusters interacting
with one another.  
The prospects of finding X-ray gas originating within
filaments (at least at redshifts up to $z\sim0.1$)
is improving, however, due to the advent
of satellites such as {\it XMM-Newton} (Pierre, Bryan
\& Gastaud 2000).

In the absence of X-ray data, other methods can also be used.
Pimbblet \& Drinkwater (2004) find a significant overdensity of 
galaxies between the close clusters Abell~1079 and Abell~1084
by utilizing a statistical background correction technique.  
Both Dietrich et al.\ (2004) and Gray et al.\ (2002) 
analyze the region between close cluster pairs using weak
gravitational lensing to infer the existence of inter-cluster
FOGs.
Spectroscopic observations also lead to concrete 
filamentary detections (e.g.\ 
Doroshkevich et al.\ 1996;
Doroshkevich et al.\ 2001;
Pimbblet, Drinkwater \& Hawkrigg 2004; 
Ebeling, Barrett \& Donovan 2004; 
Pimbblet, Edge \& Couch 2005;
see also Kodama et al.\ 2001 who employ photometric redshifts).
In the case of less complex datasets (i.e. 2-dimensional with 
little or no colour information), FOGs can still be found 
by making use of techniques such as {\it Shapefinder} statistics
(e.g.\ Bharadwaj et al.\ 2000; Pandey \& Bharadwaj 2004);
genus statistics (e.g.\ Hoyle et al.\ 2002; Hoyle, Vogeley \& 
Gott 2002);
minimal spanning trees (Doroshkevich et al.\ 2001);
marked point processes (Stoica et al.\ 2005)
and other multiscale approaches (e.g.\ Arias-Castro et al.\ 2004 and
references therein).

This work presents a new algorithm to detect galaxy 
filaments in such low-complexity datasets by utilizing
galaxy alignments.  In section~2 we present the algorithm
and the reasoning behind it.  We evaluate our algorithm in section~3
by testing it
on the 2dF Galaxy Redshift Survey (Colless et al.\ 2001).
Our results are discussed in section~4 and we summarize our major
conclusions in section~5.

\section{The Algorithm}

The relative alignments and orientations of galaxies and clusters 
of galaxies has a long history.  Binggeli (1982)
and Struble (1990) find that the major axis of 
galaxy clusters are generally aligned exceptionally 
well with their first-ranked (usually a cD-type) galaxy 
and that close ($<30 h^{-1}$ Mpc) cluster pairs generally `point 
to each other'; re-affirming the earlier work of 
Carter \& Metcalfe (1980).  Later work confirmed these 
results (e.g.\ 
West \& Blakeslee 2000; Kitzbichler \& Saurer 2003;
Pereira \& Kuhn 2004; see also Cabanela \& Aldering 1998)
and indicated that alignment effects between clusters
can range up to tens of Mpc (e.g.\ 
Lambas et al.\ 1990; West 1994; Plionis 1994).
Further, Fuller, West \& Bridges (1999) find that
the alignment effect between first-ranked galaxies and their
host cluster, and between near cluster neighbours, is not
restricted to only rich clusters, but also extends to much poorer
clusters and galaxy groups as well.  

Within clusters themselves, substructure is also found to align 
well with cluster orientation and with larger-scale 
filaments that feed cluster's growth (Plionis \& Basilakos
2002; West, Jones \& Forman 1995; see also Plionis et al.\ 2003;
Novikov et al.\ 1999; Kitzbichler \& Saurer 2003).
This scenario is (naturally) well-supported in
$\Lambda$CDM hierarchical structure modelling 
(e.g.\ West, Villumsen, \& Dekel 1991;
van Haarlem \& van de Weygaert 1993; West 1994; 
Bond, Kofman, \& Pogosyan 1996;
Dubinski 1998; Splinter et al.\ 1997; Tormen 1997;
Hatton \& Ninin 2001; Faltenbacher et al.\ 2002;
Knebe et al.\ 2004; Hopkins, Bahcall \& Bode 2005
amongst others) where filamentary structure funnels material
along preferred directions toward clusters.  
If galaxy alignment does tend to follow the orientation of
clusters and the filaments that feed them
(and indeed their own galaxian neighbours; Mackey et al.\ 2002), 
then we can potentially 
make use of this fact in order to better detect and constrain
the locations of inter-cluster FOGs.  We note, however, that 
this assumption is unlikely to work in the case of very isolated 
clusters as violent relaxation will likely have eliminated
any primordial alignments (Plionis et al.\ 2003; Coutts 1996;
see also Quinn \& Binney 1992; Lee 2004).  Plionis et al.\ (2003) use
this fact to distinguish between dynamically active,
young and still accreting clusters
(ones that have significant alignment between the cluster axis
and constituent galaxies other than the first-ranked one) 
and inactive ones.
Although Knebe et al.\ (2004) do not fully concur, they point out
that filaments are indeed well-aligned with the halo that they feed.

Our algorithm broadly follows the procedure outlined
by Plionis et al.\ (2003; see also Struble \& Peebles
1985).  Firstly, we select a (circular; square) region
(with radius $r$; of side $l$) of sky that is of interest to us
(e.g.\ a region that contains a galaxy cluster)
and extract from it all galaxies, $N$,
with known position angles, $\theta_i$ ($1 \leq i \leq N$), 
relative to some 
cardinal vector (say East--West, for instance).  
The exact choice of which galaxies to use is explored in more
detail in section~\ref{sec:bias}.
We then compare these angles to a proposed FOG angle, 
$\theta_f$ ($0 < \theta_f \leq 180$),
again measured from the same cardinal direction:

\begin{equation}
\phi_{i,f} \equiv \bracevert \theta_i - \theta_f \bracevert
\end{equation}

For an isotropic distribution, $<$$\phi_{i,f}$$>$$\approx 45$
degrees.  Hence, from the values of $\phi_{i,f}$ we can quantify the 
degree of anisotropy by following Struble \& Peebles (1985):

\begin{equation}
\delta = \sum_i \frac{\phi_{i,f}}{N} - 45
\end{equation}

which has a standard deviation thus:

\begin{equation}
\sigma = \frac{90}{12 N ^{1/2}} 
\end{equation}

We interpret the resultant value of $\delta$ according to
Table~\ref{tab:interpret}.  Since our proposed filament angle,
$\theta_f$, may likely be wrong, we proceed to
compute $\delta$ for the whole range
of $0 < \theta_f \leq 180$ 
and find a value of $\theta_f$ that 
minimizes $\delta$ (i.e.\ we find the filament angle that
aligns best with all $\theta_i$ in our particular
region of sky; Table~\ref{tab:interpret}).

Assuming that there is a $\theta_f$ vector that minimizes
$\delta$, we can proceed in an iterative fashion
by choosing a new region of sky in the direction indicated
in order to trace out any (inter-cluster) FOG present.

%
% TABLE.  delta value interpretations
%
\begin{table*}
\begin{center}
\caption{Interpretations 
of $\delta$ values.
\hfil}
\begin{tabular}{cl}
\noalign{\medskip}
\hline
Situation & Interpretation \\
\hline
$\delta \approx 0$ & An isotropic distribution. \\
$\delta <0$        & Non-isotropic distribution.  The galaxy angles align with the proposed filament angle. \\
$\delta >0$        & Non-isotropic distribution.  The galaxy angles misalign with the proposed filament angle. \\
\hline
\noalign{\smallskip}
\end{tabular}
  \label{tab:interpret}
\end{center}
\end{table*}

\section{Testing the Algorithm}

For convenience, we elect to test out our algorithm on the
2dFGRS catalogue (Colless et al.\ 2001) since it has
already been searched visually for galaxy filaments by
Pimbblet, Drinkwater \& Hawkrigg (2004).  
The observations made by 2dFGRS are summarized by Colless et
al.\ (2001) and here we only recount the pertinent detail.
Briefly, the input catalogue for 2dFGRS is
the APM survey of Maddox et al.\ (1990a,b).
Targets for 2dF spectroscopy are selected\footnote{The selection
process does introduce some incompleteness (Pimbblet et al.\ 2001;
Cross et al.\ 2004).} to be
brighter than an extinction-corrected magnitude 
limit of $b_J=19.45$ within three strips of the APM survey
(NGP, SGP and random fields) covering an area in excess
of $1500$ square degrees.
Subsequently, quality (quality$\geq3$; see Colless et al.\ 2001) 
redshifts for 221414 galaxies have been published as part of 
the 2dFGRS FDR.

\subsection{Biases}\label{sec:bias}

\begin{figure}
\centerline{\psfig{file=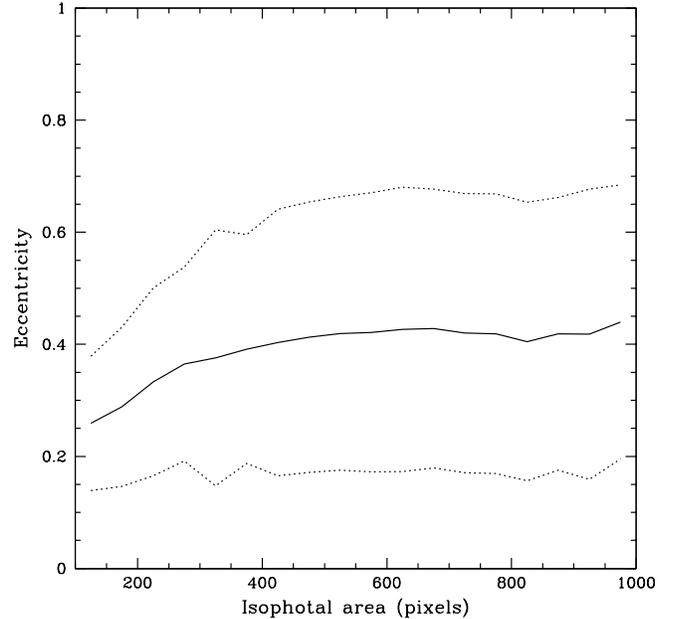,angle=0,width=3.5in}}
  \caption{Galaxy eccentricity as a function of galaxy isophotal area
for all galaxies with $b_J < 19.0$ in the NGP (the SGP follows 
a similar trend).  
The solid line denotes a running 
mean eccentricity and the dotted lines are $1\sigma$ errors on
the mean.  
For galaxies with an isophotal area less than
about 400 pixels, there is a systematic dip in eccentricity. Note that
the plot is limited to galaxies with isophotal areas less than 1000 pixels.
Beyond this range, the running mean remains near-constant about an
eccentricity of approximately $0.43$.
}
\label{fig:ellip}
\end{figure}

For 2dFGRS galaxies selected from the APM, one should expect
that there would be no systematic bias with galaxy size.  
As found in a different sample by Plionis et al.\ (2003),
galaxy size does become biased for smaller values of
galaxy eccentricity.  Figure~\ref{fig:ellip} displays 
the relationship between isophotal galaxy size and
eccentricity (as measured in the APM survey; see Maddox et al.\ 1990a
for in depth descriptions of these quantities) for bright
galaxies with $b_J<19.0$
(some $3\sigma$ away from 2dFGRS's magnitude limit of $b_J=19.45$;
Pimbblet et al.\ 2001; Colless et al.\ 2001).
At low values of isophotal area ($<400$ pixels), there is
a clear dip in the mean eccentricity.  
This bias is likely to be due to attempting to determine
galaxy eccentricity from a limited, small number of 
pixels (see also Plionis et al.\ 2003).
From herein, therefore, we only utilize galaxies with an
isophotal area of greater than 400 pixels.  This area is
determined by making trial-and-error cuts in isophotal area
and running the position angle test (described below) on
the resultant galaxies.  
In order to use only galaxies that have distinct elongations,
we also limit our selection to galaxies with an eccentricity
$> 0.05$.

\begin{figure}
\vspace*{-1.2in}
\centerline{\psfig{file=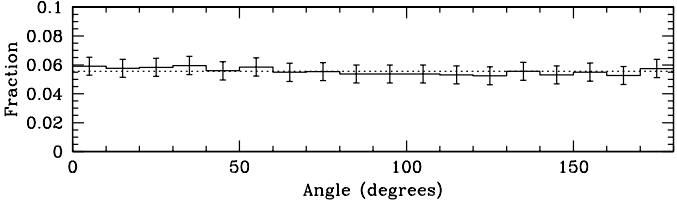,angle=0,width=3.75in}}
\vspace*{-1.2in}
  \caption{Histogram of the position angles, $\theta_i$, of
galaxies selected from 2dFGRS.  Errorbars are simple Poissonian ones.
A cursory inspection of this data suggests that the distribution
of $\theta_i$ is approximately flat: all of the bins
are within $1\sigma$ of the 
expected mean (denoted by the horizontal dotted line).
}
\label{fig:anghist}
\end{figure}

Prior to examining 2dFGRS for filaments, it is also necessary
to confirm that the position angles, $\theta_i$, of galaxies within it 
are free from any contaminating biases.  
Figure~\ref{fig:anghist} displays a histogram of position angles
of all 2dFGRS galaxies.  The distribution is approximately flat, 
containing no bin that is more than $1\sigma$ away from the
expected mean value.

%
%  FIGURE.  Area vs eccentricity
%
%

To better test for bias (or lack thereof) in these position 
angles, we utilize the Fourier transform of the position angles
following the prescription of
Struble \& Peebles (1985; also see Plionis et al.\ 2003):

\begin{equation}
C_n = \left(\frac{2}{N}\right)^{1/2} \sum_{1}^{N} \cos 2n\theta_i
\end{equation}

\begin{equation}
S_n = \left(\frac{2}{N}\right)^{1/2} \sum_{1}^{N} \sin 2n\theta_i
\end{equation}

where the position angle $\theta_i$ runs from 1 to $N$ and $n$
is an integer $\geq 1$.  Assuming that the position angles are
randomly distributed, the average values of $S_n$ and $C_n$ should
be zero with a standard deviation of 1, assuming $N > > 1$ for a
Gaussian distribution (Struble \& Peebles 1985)\footnote{The actual
purpose of using the Fourier transforms is to examine if the galaxies
have a particular preferred direction. To illuminate this point, consider
the $C_1$ component -- a test for deviation from 90 and 180 degrees 
directions.}.  
The fundamental and the first few harmonics for the sample is
noted in Table~\ref{tab:harmonics}.  The individual values 
show no significant bias (i.e.\ a value $>>3$)
and the distribution is consistent with a $N(0,1)$ Gaussian
under a standard KS test;
thus we consider that these values are consistent with isotropy.  
Without the cut in galaxy isophotal
area, these values are found to become significant
indicating a distinct bias to a preferred direction(s).

%
%  FIGURE.  Position angle histogram
%
%

%
% TABLE.  Harmonics of Cn and Sn.
%
\begin{table}
\begin{center}
\caption{Fourier transform of the distribution of
position angles, $\theta_i$.  
We consider that these values are consistent 
with isotropy.
\hfil}
\begin{tabular}{ccc}
\noalign{\medskip}
\hline
$n$ & $C_n$ & $S_n$ \\
\hline
1 & 2.6 & -0.3 \\
2 & 1.3 & -1.3 \\
3 & 0.6 & -0.6 \\
4 & -0.9 & 0.5 \\
5 & -0.3 & 1.2 \\
6 & 1.1 & -1.3 \\
\hline
\noalign{\smallskip}
\end{tabular}
  \label{tab:harmonics}
\end{center}
\end{table}

\subsection{Case study: Abell~1651}

We start our evaluation by testing the algorithm on the (known)
case of Abell~1651.  Work by Pimbblet, Drinkwater \& Hawkrigg (2004)
found that Abell~1651 is connected to Abell~1663 (a nearby cluster)
by a Type I filament (i.e.\ a straight filament).

\begin{figure}
\centerline{\psfig{file=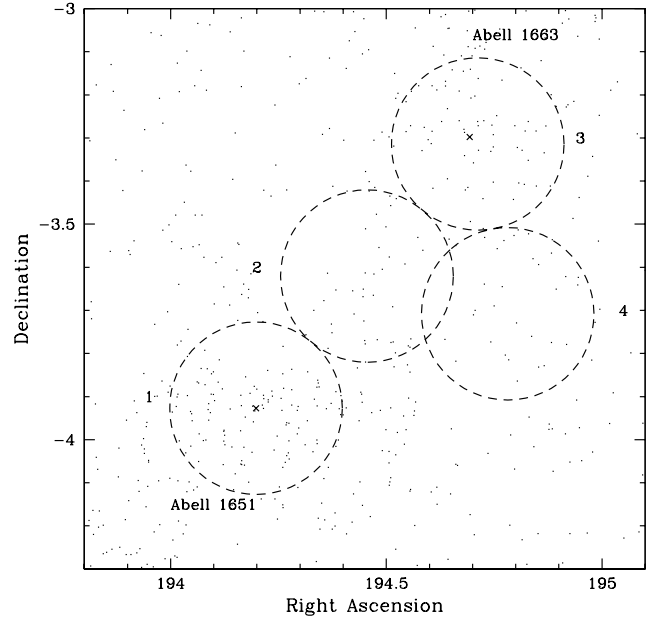,angle=0,width=3.5in}}
  \caption{Area of 2dFGRS investigated.  
The crosses denote the approximate
positions of Abell~1651 and Abell~1663 whilst the dashed circles are
the positions of the $r=0.2$ degree circular regions investigated using 
our algorithm.  The numbers show the order in which the $r=0.2$
circular regions are analyzed, starting at Abell~1651.}
\label{fig:path}
\end{figure}

\begin{figure}
\centerline{\psfig{file=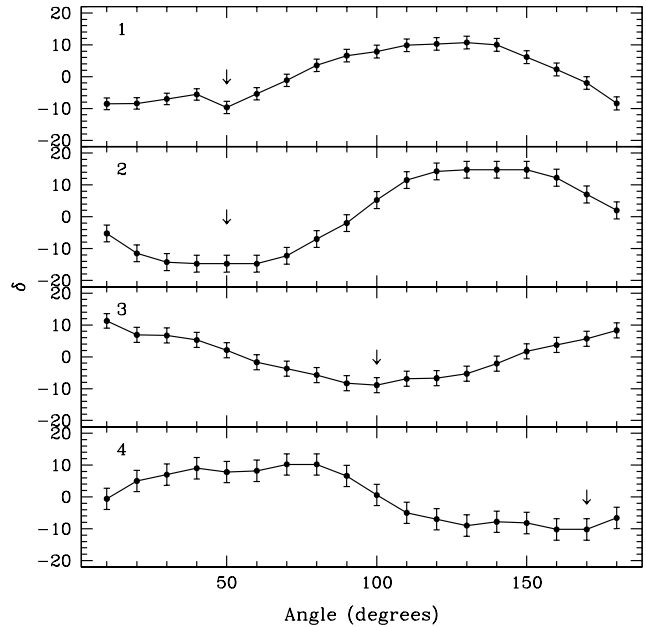,angle=0,width=3.5in}}
  \caption{The $\delta$ values as a function of $\theta_f$ for
each of the $r=0.2$ degree circular regions identified from 
Figure~\ref{fig:path}. The downward pointing arrows show
the minimum value of $\delta$ for each region.  Errors are
from equation~(3).}
\label{fig:delta}
\end{figure}

Firstly, we extract from 2dFGRS a region of galaxies contained within
$r=0.2$ degrees from the cluster centre of Abell~1651.
Using these galaxies, we search for a correlation in direction by 
finding the angle $\theta_f$ that minimizes $\delta$.  
We then repeat this process for a new region of sky in the direction
indicated by the previous step (note that as this result shows two possible
vectors, 180 degrees apart, we always choose the vector pointing
closest to the next nearest cluster).
The results of this analysis are displayed in Figures~\ref{fig:path} 
and~\ref{fig:delta} which show the positions of the circular regions
used and the values of $\delta$ against $\theta_f$ respectively.

%
%  FIGURE.  Path: A1651
%
%

%
%  FIGURE.  delta values: A1651
%
%

Going from Abell~1651 to Abell~1663 is relatively easy: the first two
steps produce exactly the same result ($\delta_{min} = 50$ degrees),
making for a very straight filamentary connection between the two
clusters.
In reverse, however, the clusters are still connected, but the 
connection becomes more curved.  Step (3) (located nearly on top
of Abell~1663) results in 
$\delta_{min} = 100$ degrees, which takes the path of the filament
slightly away from the original straight one.  
This is likely due to contamination from
galaxies on the opposite side of Abell~1663: taking only those
galaxies in the semi-circle of step (3) 
closest to Abell~1651 would
have yielded a position for step (4) that is much closer to step (2). 

Nonetheless, the situation is rectified in the `real'
step (4) where $\delta_{min} = 170$ degrees.
This shows that the next step
would land the path back almost on top of step (2).
Indeed, Pimbblet, Drinkwater \& Hawkrigg (2004; PDH)
find that the connection between Abell~1651 and Abell~1663 is
reasonably straight: a Type~I filament in their nomenclature.  
This is certainly the case where one starts at Abell~1651.  
By starting at Abell~1663, however, would the resultant  
curvature have affected the (visual typing) result by PDH?  
In all likelihood, no.  
The distance of step (4) away from the inter-cluster axis
is just less than about
0.4 degrees (Figure~\ref{fig:path}).  At the mean 
redshift of the clusters ($z\approx0.084$; 
Pimbblet et al.\ 2001; Pimbblet 2001)
this separation equates to no more than 1.5 Mpc $h^{-1}_{100}$, 
i.e.\ very close to the inter-cluster axis and plenty 
smaller than the three-dimensional inter-cluster separation 
of $\approx 9$ Mpc $h^{-1}_{100}$ (PDH).

We note that even if the algorithm is able to find a filament
between clusters A and B (by starting at cluster A), it
does not necessarily follow that it will be able to find 
the same filament by starting at cluster B instead.  
This does not mean that the algorithm does not work.
Both clusters A and B could be connected to multiple
filaments.  Indeed, the expected number of filaments 
connected to a given cluster
scales well with the cluster mass (Colberg, Krughoff \& 
Connolly 2004; PDH).
The algorithm will preferentially find the neighbouring cluster
or filament from whose direction the last clump of 
material fell in from.  
We also note that the algorithm will likely become confused
if there is close alignment (i.e.\ superposition)
of structures in the $z$ direction.

\subsection{Robustness}

The 2dFGRS catalogue is known to be approximately 10 to 
20 per cent incomplete at all magnitudes 
(Pimbblet et al.\ 2001; Cross et al.\ 2004).
The `missing galaxy' population is not preferentially situated 
near cluster cores (Pimbblet et al.\ 2001) and can therefore
be simulated as a purely additional
random galaxy sample that follows the clustering pattern
and has random orientations, $\theta_i$.
Therefore to test if we can recover the filament signal in the
presence of, what is essentially, increased noise, we repeat our
experiment by adding in 15 per cent more galaxies in 100 Monte Carlo 
realizations.

The median value of $\delta_{min}$ out of the 100 realizations
for steps (1) and (2) is then
$50 \pm 39$ and $50 \pm 14$ degrees respectively.
Since step (2) is directly inbetween the clusters, there is only
a small amount of error on the median value.  
Step (1), however, is located at a cluster and
the larger error on the average $\delta_{min}$ value is due to 
galaxies at the opposite end
of the cluster to the filament, similar to what is seen at step (3),
above.  Indeed, the majority of the Monte Carlo realizations that do not
result in $\delta_{min}=50$ for step (1) generate $\delta_{min}=180$.
From Figure~\ref{fig:delta}, it can be seen that $\delta=180$ for step
(1) is also a local minima.
It is likely, therefore that Abell~1651 has multiple filaments
or significant substructure (see Plionis \& Basilakos 2002)
falling in along the $\delta=180$ vector.

\section{Discussion}
\begin{figure}
\centerline{\psfig{file=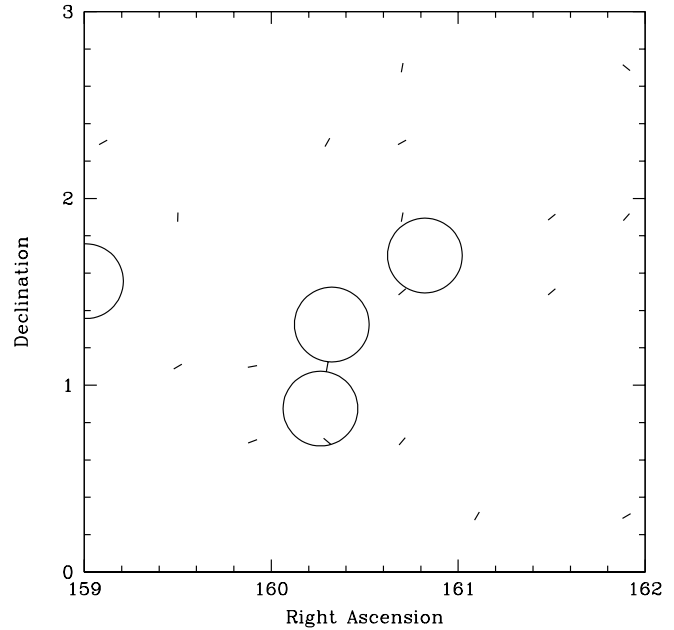,angle=0,width=3.5in}}
  \caption{The algorithm as applied to 
the Leo-Sextans supercluster region of the NGP.
Each line displays the vector corresponding to $\delta_{min}$
for each bin with a significant detection:
$(\delta_{min} / \sigma) > 4.0$.  The open circles denote the locations
of known galaxy clusters taken from De Propris et al.\ (2002).
Close cluster pairs display vectors pointing toward one another.
}
\label{fig:ngp}
\end{figure}

To better examine any filaments present in 2dFGRS we now 
apply the algorithm to the entire NGP 2dFGRS dataset. 
This is accomplished by 
dividing the 2dFGRS NGP data up into squares of side $l=0.2$ degrees
and applying the algorithm to each square sequentially.
Figure~\ref{fig:ngp} displays the result of this in the region
of the Leo-Sextans supercluster.
Close cluster pairs,
especially in regions of high cluster density (i.e.\ superclusters),
display obvious (direct) inter-connections, but galaxy clusters 
at larger (say $>>15$h$^{\rm -1}$ Mpc or $>>$ few degrees) separations 
seldom appear to do so.

%
%   FIGURE.  NGP filaments - zoom-in of Leo-Sextans. 
%
%

Clearly the arbitrary choice of the centres of the $l=0.2$ degree
squares can affect our results significantly.  Therefore, we 
repeat the analysis that we performed on Abell~1651 \& Abell~1663 
(see above) on other (close) cluster pairs identified previously by PDH.
Comparing to PHD, we conclude that our algorithm 
has an efficiency of $\sim 75$ per cent 
at detecting filaments at inter-cluster 
separations of up to 15 h$^{\rm -1}$ Mpc.
Morphologically, these filaments are mostly
straight with some slightly curved ones
(i.e.\ Type~I and some Type II filaments; PDH).
At larger inter-cluster separations, however, the detection 
efficiency rapidly decreases to zero for all filament types
regardless of whether they are morphologically Type~I or II.

Further, we find that other types of filament morphology such
as `walls' or, equally, `sheets' (Type III filaments; PDH) do not 
display significantly cohesive galaxy orientations.  
This is borne out by re-analyzing the recent
work of Pimbblet, Edge \& Couch (2005) who have
discovered a large scale wall in the direction of Abell~22.
We find that the galaxies belonging to their 
Type~III filament do not display any 
preferred (global) galaxy orientation.
We reach the same conclusion by re-analyzing the Type~III
and IV (`clouds') filaments found by PDH.
With more irregular filaments (Type~V; PDH), there are sometimes
($\sim$ 10--25 percent of all bone-fide Type~V filaments)
significant galaxy orientations; particularly in the cases where
galaxy clusters are joined by close multiple filaments.  
In highly irregular and lumpy Type~V filaments, the orientations
are statistically consistent with isotropy.

Why do only relatively straight and short 
inter-cluster filaments of galaxies
possess significant non-isotropic distributions of 
galaxy orientation angles with respect to the filament in
which they reside?
Consider a brief thought experiment utilizing the laminar flow model (LFM)
outlined by Kitzbichler \& Saurer (2003; see also Aubert, Pichon \& 
Colombi 2004).  We know that filaments of galaxies
can be thought of as funnels that direct material onto clusters
(e.g.\ Plionis \& Basilakos 2002; Knebe et al.\ 2004).
In the direction that is perpendicular to the funnel's direction
(the infall direction), matter is generally becoming more dense and
coalescing.  Conversely, along the vector parallel to the infall direction
matter will be elongated by tidal force.

Consider the case of a single, isolated, straight filament joining 
together two close galaxy clusters.  The LFM succinctly suggests 
that the galaxies in the filament will elongate along the 
filament's direction as they move to either
end (i.e.\ the clusters) of the filament.
If separated by a greater distance, those galaxies that start out
near to the clusters will be the first ones to elongate 
significantly as time passes and the elongation will become more
pronounced as they move closer to the gravitational potential
of the cluster.  
Indeed, if the galaxies only move by up to $\sim 10$h$^{\rm -1}$ 
Mpc since their creation (Coles, priv.\ comm.)
it is little wonder that galaxies beyond $>>10$h$^{\rm -1}$ Mpc
from a cluster have had much chance to elongate and align 
with the filament, hence the longer the filaments 
(say, $>>15$h$^{\rm -1}$ Mpc), the less likely 
they are to be detected by this method.

The situation is more complicated in the case of curved filaments.
In general, the curving of such filaments is known to be 
toward a relatively far-removed tertiary mass (Pimbblet, 
Drinkwater \& Hawkrigg 2004; Colberg, Krughoff \& Connolly 2004).
The presence of such a tertiary mass (e.g.\ another cluster) 
would have the effect of `confusing' the galaxy orientations; 
particularly of those that start out distant from the two primary 
clusters.  With other filament morphologies (and indeed, branching
of them), the gravitational 
attraction of the filament itself will become much more 
important, further mixing the position angles of constituent galaxies.

\section{Conclusions}

This work presents a new algorithm for detecting filaments of
galaxies between clusters based upon the assumption that their 
constituent galaxies will possess orientations
that are not isotropic in nature.  
Our major findings are as follows:

\begin{itemize}

\item The algorithm will preferentially find the filament
from whose direction the last clump of material fell in 
from.  This means that one may not recover the exact same filament
between clusters A and B by starting at cluster A versus starting
at cluster B.

\item The algorithm works well on detecting the
presence of straight Type~I
inter-cluster filaments of galaxies
separated by short distances (up to 15 h$^{\rm -1}$ Mpc).  
Up to $\sim 75$ per cent of
such filaments are easily recovered using this algorithm.

\item Longer Type~I and many Type~II filaments do not display 
the same non-isotropic distribution of galaxy orientations that 
their shorter counterparts do.

\item The orientations of galaxies belonging to Type~III and 
IV inter-cluster galaxy filaments are statistically consistent 
with isotropy.

\item We suggest that low detectability of long filaments and 
more irregular filaments can be explained by considering 
a basic laminar flow model: galaxies in straight, isolated filaments
are more likely to become elongated toward the clusters that they
connect than galaxies in filaments with more complex morphologies.

\end{itemize}

It would be interesting to investigate the star-formation rate
along these filaments (say by using the $\eta$ parameter suggested
by 2dFGRS; see Madgwick et al.\ 2002) to see if galaxies that are 
aligned well with a filament have a vastly different value to those
that are not (Plionis et al.\ 2003).  It should also be the case that
galaxies near the centre of filaments are much younger (more blue and 
spiral-like) than galaxies toward a terminus (i.e.\ a cluster).

This algorithm could also be used to find potential filaments
that are hiding in the Zone of Avoidance (providing, of course, that
individual galaxy ellipticities can be measured accurately there).
For example, it would be a good, independent
test of the suspected filament found by Kraan-Korteweg, Woudt \& 
Henning (1997) connecting the Hydra and Antlia clusters.
Lee (2004), however, cautions that on its own evidence favouring 
`coherent orientation of 
galaxies embedded in a sheet should not be taken as identical 
to the existence of the sheet itself'.

Lastly, we note that the issue of detecting FOGs in redshift surveys can
also be thought of as a `join-the-dots' type problem
(Arias-Castro et al.\ 2004; see also Stoica et al.\ 2005).  
Using galaxy orientations 
would readily transform this into the vectorized problem of 
`join-the-darts' (Arias-Castro et al.\ 2004) and it will
be interesting to to see, in the future, how the results of these 
algorithms compare to the one presented here.

\section*{Acknowledgments}
This work has benefitted enormously from 
conversations with participants attending the November 2004 Multiscale
Geometric Analysis Workshop IV, held in Los Angeles and 
I wish to warmly thank Jean-Luc Starck for inviting me to participate.

I also want to express my gratitude to the referee, 
J\"{o}rg Colberg, for a prompt and useful report 
that has improved this work.

KAP was supported by an 
EPSA University of Queensland Research Fellowship and 
a UQRSF grant throughout the course of this work.

\end{document}